\documentclass[conference]{IEEEtran}
\IEEEoverridecommandlockouts

\usepackage{cite}
\usepackage{subcaption}
\usepackage{amsmath,amssymb,amsfonts}
\usepackage{algorithmic}
\usepackage{graphicx}
\usepackage{subcaption}
\usepackage{multirow}   
\usepackage{textcomp}
\usepackage{xcolor}
\usepackage{url}
\usepackage{listings}
\usepackage{framed}
\usepackage{booktabs}
\usepackage{longtable}

\def\BibTeX{{\rm B\kern-.05em{\sc i\kern-.025em b}\kern-.08em
    T\kern-.1667em\lower.7ex\hbox{E}\kern-.125emX}}

\begin{document}

\title{LLM Contribution Summarization in Software Projects}

\author{\IEEEauthorblockN{1\textsuperscript{st} Rafael Corsi Ferrão}
\IEEEauthorblockA{\textit{Insper} \\
São Paulo, Brazil \\
{rafael.corsi@insper.edu.br}}
\and
\IEEEauthorblockN{1\textsuperscript{nd} Fabio de Miranda*}
\IEEEauthorblockA{\textit{Insper} \\
São Paulo, Brazil \\
{fabiomiranda@insper.edu.br} 
\footnote{Equal contribution.}}
\and
\IEEEauthorblockN{2\textsuperscript{nd} Diego Pavan Soler}
\IEEEauthorblockA{\textit{Insper} \\
São Paulo, Brazil \\
{diegops@insper.edu.br}}
}

\maketitle

\begin{abstract}
This full paper in innovative practice provides an automated tool to summarize individual code contributions in project-based courses with external clients. Real industry projects offer valuable learning opportunities by immersing students in authentic problems defined by external clients. However, the open-ended and highly variable scope of these projects makes it challenging for instructors and teaching assistants to provide timely and detailed feedback. This paper addresses the need for an automated and objective approach to evaluate individual contributions within team projects. In this paper, we present a tool that leverages a large language model (LLM) to automatically summarize code contributions extracted from version control repositories. The tool preprocesses and structures repository data, and uses PyDriller to isolate individual contributions. Its uniqueness lies in the combination of LLM prompt engineering with automated repository analysis, thus reducing the manual grading burden while providing regular and informative updates. The tool was assessed over two semesters during a three-week, full-time software development sprint involving 65 students. Weekly summaries were provided to teams, and both student and faculty feedback indicated the tool’s overall usefulness in informing grading and guidance. The tool reports, in large proportion, activities
that were in fact performed by the student, with some failure to detect students contribution. The summaries were considered by the instructors as a useful potential tool to keep up with the projects.
\end{abstract}


{\footnotesize \textsuperscript{}
\thanks{This work has been submitted to the IEEE for possible publication. Copyright may be transferred without notice, after which this version may no longer be accessible}

\thanks{*Equal contribution between Rafael Corsi Ferrao and Fabio de Miranda}}

\section{Introduction}\label{Introduction}

Longer undergraduate programming projects like capstones, hack\-ath\-ons, or sprints are often characterized by team collaboration and intense student dedication. Frequently, the client is an external entity, such as a company representative, and in these scenarios, the student team is responsible for structuring and managing the project. These types of projects can address a common deficiency in undergraduate training related to soft skills, such as communication and teamwork, often highlighted in the literature. However, the intensive nature of these projects can overload the instructor due to the amount of work involved in keeping track of multiple student groups and individuals' work. Consequently, formative feedback may suffer, and summative feedback often relies on presentations or other superficial evidence.

Large Language Models (LLMs), based on transformers, have been applied to various activities related to programming, natural language processing, and data analysis. These models combine extensive training data with a smaller context window to produce outputs that are frequently coherent. LLMs have proven potential in education \cite{lo_what_2023}, for instance, in grading open-text responses \cite{feldt_ways_2018}, and assisting students in resolving programming errors without providing direct answers \cite{phung_generating_2023}.

This article reports on the experience of applying LLMs to monitor the progress of student teams during a course that consists of an intensive 3-week code development sprint. In this context, the requirements came from a real representative from the industry, and the design of the software was entirely up to the students. So, the instructor couldn't impose a common structure on the projects to make monitoring or assessment more feasible. 

The expectation is that the LLM can be used to enable a timely and individualized summary of work accomplished on intensive, industry-oriented software projects. The summary provided to each student's work favors the self-regulation of group work, since it was available to instructors and all the students. This tool intersects with the subjects of collaborative teamwork, technology-assisted work, and the use of AI in education. 

\section{Literature Review}\label{literature-review}


In recent times, there has been considerable interest in applying LLMs in education, primarily due to the impressive results demonstrated by ChatGPT, which generates responses that are consistent and systematic enough to be considered useful. Lo \cite{lo_what_2023} reviewed the literature on the impact of ChatGPT in education and identified various application examples, including generating customized assessment items, feedback, and guidance for open-ended activities such as group essays \cite{kasneci_chatgpt_2023, baidoo-anu_education_2023}, simulating a peer in discussion groups \cite{gilson_how_2023}, and facilitating debates while providing personalized feedback.

Dehbozorgi \cite{dehbozorgi_llm-based_2024} applied LLMs to provide feedback on formative questions developed by students about the class content. The LLM, having access to the course materials, was able to assess the relevance and alignment of these questions with the course topics. Rudolph \cite{rudolph_chatgpt_2023} identified that LLMs can be used to personalize student support systems through \textit{Intelligent Tutoring Systems} (ITS), providing original cases for discussion tailored to each group of students. Provided that ethical issues and data privacy concerns are safeguarded, LLMs could enable adaptive personal tutors, which can consider a student's history of actions, personal, and emotional states, thereby offering personalized feedback and suggestions. Cope \cite{cope_artificial_2021} mentioned that one of the greatest potentials for transforming education lies in using LLMs to provide more consistent and rich formative feedback.

The fields of Computing and Software Engineering have significant potential to be influenced by LLMs, particularly because the artifacts produced in these professions are digital language products, which are the niche where LLMs excel. Kirova \cite{kirova_software_2024} emphasized the need to rethink the teaching of Software Engineering in an LLM context, recognizing their potential for writing code, testing, and general automation, while also understanding how they work to identify risks such as intellectual property and security concerns.

\subsection{Feedback on source code repos}

Projects created for the industry serve as a valuable complement to theoretical student training \cite{robles_development_2022, radermacher_investigating_2014}. Teamwork is an important skill on the job market \cite{ruf_communication_2009, ettington_facilitating_2002}. However, it is necessary to provide proper follow-up to ensure that the team work experience is both educational and productive. It is widely accepted that simply grouping students and expecting them to learn teamwork unsupervised is insufficient \cite{bacon_lessons_1999}.

Monitoring and providing feedback during the development of a software project are fundamental. In the context of capstones or intensive sprints such as the one in this report, it is impractical for the professor to closely track the progress of each group. This literature review will address previous attempts to monitor academic or non-academic software projects using Git repositories.

Several authors \cite{robles_development_2022, radermacher_investigating_2014} point out that the ability to work in a team is one of the most important skills for the job market. Group projects are also an opportunity to develop communication, another skill commonly pointed out as deficient in engineering and computing graduates \cite{ruf_communication_2009}.

However, students often encounter negative experiences when working in teams within a school context \cite{feichtner_why_1984, falkner_collaborative_2013}. On one hand, they dislike the uneven levels of contributions among group members; on the other hand, they tend to avoid conflicts to address this issue and are reluctant to expose peers who are not contributing effectively.

It is important to note that not all group work effectively prepares students for teamwork in the professional world \cite{ettington_facilitating_2002}. Group projects are intended to simulate professional teams, where deliverables should require interactions among members and benefit from them, making the gains from these interactions visible to the students. However, students often organize themselves in ways that minimize interdependence among group members, which results in missing learning skills that are transferable to the professional world.

Common didactic strategies to improve teamwork include mutual evaluations through questionnaires based on rubrics that exemplify good teamwork within the group, such as CATME \cite{ohland_comprehensive_2012-1}, a tool used in many universities. However, individuals in teams are often reluctant to provide incisive feedback to peers \cite{willcoxson_its_2006} until a crisis occurs, which can affect the reliability of records captured by CATME. Additionally, feedback based on questionnaires cannot be provided continuously. These considerations highlight the need for an objective metric to quantify the work and collaboration of team members.

In the software field, version control systems have been well-established for decades, with Git being the most popular. This widespread adoption allows us to analyze the contributions of team members to project repositories, aiming to understand the distribution of work and collaboration within the team.

Unlike questionnaires filled out by team members, repositories and groupware can provide continuous metrics. Tarmazdi \cite{tarmazdi_using_2015} proposed a teamwork panel where the team's communication data underwent sentiment analysis to understand the team's emotional state and the roles played by each member. Additionally, the GitCanary tool \cite{sandee_gitcanary_2020} offers quantitative productivity metrics such as proportion and complexity of code committed by each team member, enabling real-time feedback on those metrics to students.

Gousios \cite{gousios_measuring_2008} demonstrated that it is possible to measure developer contributions beyond lines of code, including commit messages, bug reports, and wikis. Lima \cite{lima_assessing_2015} validated the complexity of commits and the volume of contributions as metrics indicative of positive developer contributions.

Bufardi \cite{buffardi_assessing_2020} studied the Git repositories of student teams and found a strong positive correlation between contributions to the repository and peer evaluations. He analyzed contributions from GitHub and the team's Kanban board, with the latter being manually analyzed by the instructors. The Git log evaluation was automated, focusing on quantitative metrics such as commits and lines of code. However, this analysis did not consider the content of the changed lines of code or the type of file.

Hundhausen's work \cite{hundhausen_evaluating_2021-1} analyzed projects by combining quantitative metrics and examining a random sample of code at the line-of-code level in some projects, noting that it is impractical to manually analyze all the code. Nevertheless, Hundhausen conducted a thorough analysis of the quality of the changed lines of code, employing a custom algorithm to detect changed lines of code. This allowed him to identify simple refactoring, such as code being moved rather than newly created. Additionally, Hundhausen used GitHub data to verify learning objectives related to the qualities of commits, issues, and software processes in a context of heterogeneous projects with open scope and long duration, distributed between two institutions.

\section{Context}\label{context}

This experience report takes place during a 3-week sprint at the end of the first year of the Computer Science course at INSPER \cite{igor_2024}. By this point in their studies, students had completed courses in UX/UI, programming, web development, and data science. It was performed for 2 consecutive semesters in 2024, in the first there were 28 students (3 women, 25 men), then in the second semester there were 37  students (4 women, 33 men).  In each semester, students were divided into groups of 4 to 6 students each, with the criteria of randomly grouping students of similar academic performance and ensuring that minority members were not isolated in any group. All students had previously participated in several sessions focused on group work agreements, feedback, non-violent communication, and mutual evaluation of teamwork using CATME, in an approach similar to what is described in \cite{cliquet_developing_2023}. 


The sprint is an official part of the curriculum, starting immediately after the completion of regular coursework every semester. It is a formal discipline that counts for credits, with attendance tracked for four hours each day. Outside of hours when attendance was taken, the lab was open 8am--8pm for students to continue their work, and instructors were available to answer questions. Students were encouraged to work in the lab to simulate the time commitment required in a programmer's workday. The expected workload for each student during the sprint was 30 to 40 hours per week, though strict attendance control was maintained for only 22 of these hours. The sprint was entirely practical; instructors occasionally used a whiteboard or projector for guidance or clarification, but no formal lectures were planned because the theory was already provided during the regular semester before the sprint.

The sprint has its own final grade that goes on the academic record. The group's collaboration grade considered whether students made a minimum number of commits on the frontend and backend and if they participated adequately in the CATME evaluations (details of which are not discussed in this report). To encourage honest peer evaluations, the act of completing the evaluation cycle was graded, not the grades received from peers.

During the sprint mentioned here, students met with the clients once a week to ask questions and receive feedback, and also communicated asynchronously through e-mail. Each group had a product owner, who was responsible for consolidating questions and approaching the client.

In the first semester, each group was tasked with developing a software project for a sports data analysis company. Their objective was to create a query interface for retrieving plays from a historical match database and to generate tactical field metrics related to player positioning. In the second semester, students worked on developing a portal to manage and provide access to clinical trial information for one of the largest hospitals in Brazil.

The groups had private repositories shared with the instructors and created using GitHub Classroom. Students were instructed to make commits with meaningful names, organize folders and files systematically, and, where possible, adopt a branch per issue strategy. The teams were also required to maintain an organized Scrum board with registered stories and tasks, updating it frequently. Major tasks had to be clear and divided into smaller subtasks. Besides the Product Owner, each team also had a Scrum Master.



The goal of the tool described here is to provide instructors with a brief understanding of which part of the project each student was working on, and to offer students insights to discuss the evolution and workload allocation within the group. Literature indicates that awareness of other members' progress leads to better self-regulation and negotiation within the group, enhancing teamwork and reducing free-riding behavior \cite{lin_effects_2018}. Those goals can be summarized as the following questions: 

\begin{itemize}
    \item \textbf{Q1} Does the tool provide a reliable summary of student contributions in project-based courses? 
    \item \textbf{Q2} Would such a tool be useful for instructors to keep up with all the project teams?
\end{itemize}

\section{The Summarizer Tool}\label{feedback-process}

\begin{figure*}[ht!]
  \includegraphics[width=\textwidth]{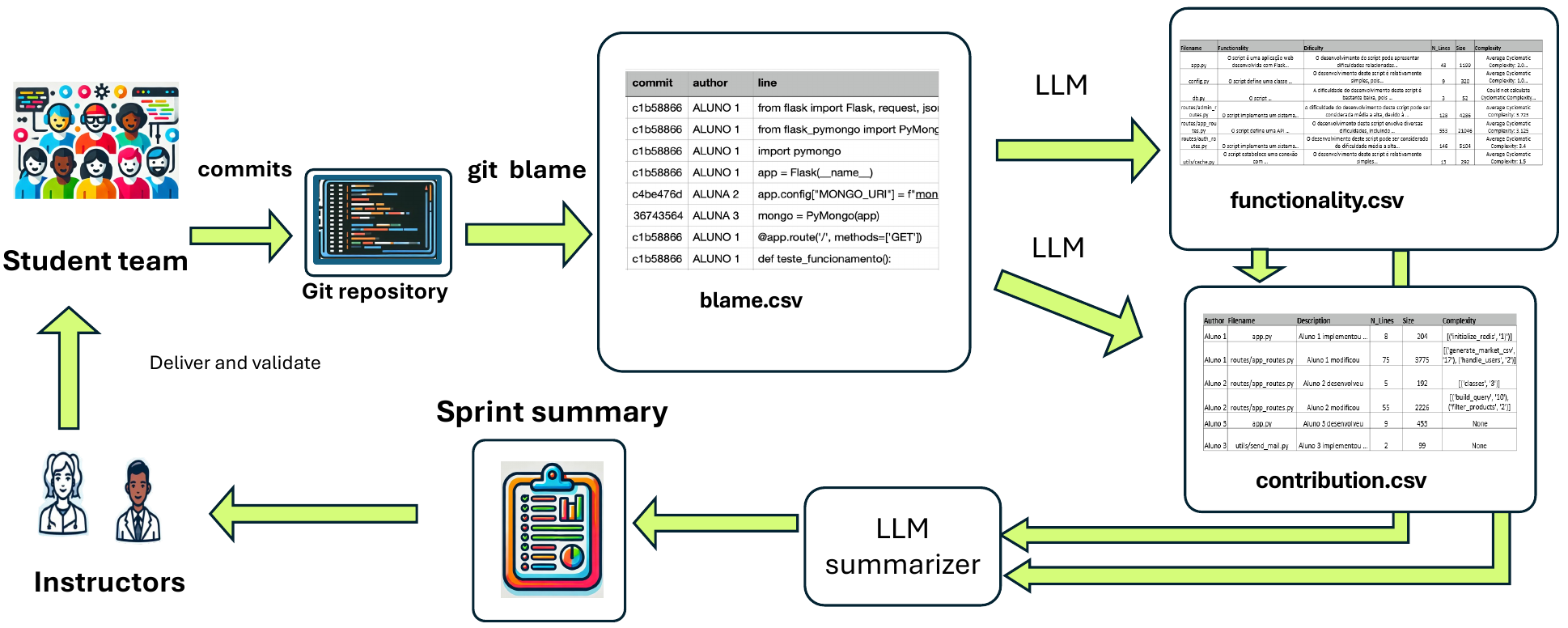}
  \caption{Process of proposed tool to generate students' code feedback using LLMs.}
  \label{fig:teaser}
\end{figure*}

This experiment aimed to create an aggregated tool capable of summarizing all the code produced by each student in a team \cite{miranda_2024}. Although the long-term goal is real-time analysis, current feedback is delivered weekly due to infrastructure and time constraints.

This tool integrates a chain of AI agents, using the \textit{GPT-4o-mini model}, to analyze and concede the changes on the group’s repository, then uses a stronger model, GPT-4o, for generating a combined, comprehensive feedback. The pipeline, illustrated in Figure~\ref{fig:teaser}, employs the OpenAI API and LangChain, complemented by a Streamlit frontend and FastAPI backend.

Initially, the process involves cloning student repositories locally. These repositories are processed with \textit{PyDriller} \cite{spadini2018pydriller} to create Git blame files, capturing each student's modifications line-by-line along with commit messages, establishing an overview of contributions. Directly feeding Git blame files to the LLM led to hallucinations and information loss; to mitigate this, we introduced a few preprocessing steps.

The preprocessing involves compressing Git blame data through a series of AI agents. The first agent analyzes each file's functionality, generating a \textit{Functionality Table} (see Tab~\ref{tab:functionality}) in \textit{.csv} format. At this stage, the LLM is prompted to read the file and provide a summary of its goals and development difficulty. Additional data collected includes file size, number of lines, and complexity metrics, cyclomatic complexity for Python scripts and Jupyter Notebooks, and HTML tag counts for HTML files.

\begin{table*}[ht!]
    \centering
        \renewcommand{\arraystretch}{1.5} 
    \begin{tabular}{|l|p{7.5cm}|p{7.5cm}|}
         \hline
         \textbf{Filename} & \textbf{Functionality} & \textbf{Difficulty}\\
         \hline
         \textit{app.py} & The script is a web application developed with Flask that configures a server, initializes connections to a MongoDB database and Redis service. It loads configurations from a \texttt{.env} file, allows requests from any origin (CORS), and registers routes for authentication, administration, and other functionalities. Additionally, it checks if the necessary collections in MongoDB exist and, if not, creates them, as well as initializes Redis by clearing all existing data. 
         & 
         The script development may present difficulties related to environment configuration, especially when dealing with environment variables and integration between different services (MongoDB and Redis). CORS implementation can be complex if not well understood, and proper structuring of routes and blueprints in Flask requires a good understanding of web application modularization. Creating unique indexes in MongoDB can also be challenging, especially if there is existing data that doesn't respect uniqueness constraints. \\
         \hline
         \textit{app\_routes.py }
         &  
         The script defines a RESTful API using Flask to manage information related to products, countries, molecules, therapeutic classes, and markets. It allows CRUD operations (create, read, update, and delete) on various MongoDB collections. The API also implements input data validation using JSON Schema, Redis caching to optimize queries, and user authentication based on tokens. Additionally, there's support for filtering products based on specific criteria and generating Excel reports. 
         &
         This script provides a comprehensive RESTful API built with Flask to manage data for products, countries, molecules, therapeutic classes, and markets. It supports full CRUD operations (create, read, update, delete) on MongoDB collections, integrates Redis for caching to optimize performance, and uses JSON Schema to validate incoming data. Secure access is enforced through token-based user authentication. In addition, the API allows advanced filtering of product records and can generate Excel reports, making it a versatile solution for complex data management needs. \\
         \hline
    \end{tabular}
    \caption{Functionality Table original generated in \textit{.csv} format by the LLM with the git blame input. Cyclomatic complexity and line count were omitted to fit the page}
    \label{tab:functionality}
\end{table*}


Next, another agent evaluates individual developer contributions to each file, generating a \textit{Contribution Table}. The LLM is prompted to describe the purpose of each developer's addition in relation to the file's functionality. Individual contributions are assessed, including the attribution of function complexities when a developer has sole authorship. An example is presented in Table~\ref{fig:contribution}, which displays reports from two students: one with a strong contribution history and \textit{Student 2}, whose contributions were identified as elementary by the tool.

Finally, we used a stronger model to synthesize and summarize a report of individual contributions, highlighting specific areas of work. For this comprehensive summary, the agent is provided with the following items and prompted to identify the contributions of each developer.

\begin{itemize}
    \item The filename, its functionality, and its complexity score.
    \item The sprint instructions
    \item A breakdown of what each developer contributed to each file, including the complexity of solo-developed functions.
    \item A set of instructions of the desired template to maintain uniformity among reports.
\end{itemize}

Optionally, the agent is tasked with classifying each developer into predefined business roles (Technical Leader, Data Engineer, Security Engineer, DevOps Engineer, Backend/Frontend Engineer and Documenter) based on their contributions. Those roles could be either Junior or Senior (e.g., Junior Backend Developer). 

\begin{table*}[ht!]
    \centering
        \renewcommand{\arraystretch}{1.5} 
    \begin{tabular}{|l|l|p{14cm}|}
         \hline
         Student & File & n\_lines \\
         \hline
         \textit{Student 1} & approutes.py &  \textit{Student 1} implemented several features in the routes/app\_routes.py file, including defining routes for market data downloads and user queries. He created the \textit{/download/} route, which allows generating and downloading an Excel file with detailed information about a specific market and its products, using data stored in a MongoDB database. For this, he developed the \textit{generate\_market\_csv} function, which fetches market and associated product data, organizes this information in a pandas DataFrame, and generates an Excel file.

         Additionally, he implemented the \textit{/users} route to return a list of users, excluding sensitive information such as passwords and emails. The functions are protected by decorators that handle database exceptions and authentication requirements. The code reflects good practices in data structuring and manipulation, while ensuring user information security.\\
         \hline
         \textit{Student 2} &  \textit{index.html} &
         \textit{Student 2} made a specific contribution to the frontend, adding a paragraph and a button to an HTML page.\\
         \hline
    \end{tabular}
\caption{The file \texttt{contribution.csv} generated by the LLM from the Git blame for a student. This includes two examples: one for \textit{Student 1}, who has a strong contribution history, and another for \textit{Student 2}, who lacks significant contributions to the project.}
\label{fig:contribution}
\end{table*}


Lastly, the LLM is also asked to generate a summary of what the team did during the sprint, and for this, it's also given the text of the description of the project and the client's high-level requirements, this appears as ``Overall contribution of the team'' in the feedback (Tab. \ref{summary-eng}). This approach leverages the LLM’s ability to provide comprehensive, role-specific insights into student contributions while making the evaluation process transparent and structured, while also reducing the chances of hallucinations and loss of information.

\begin{table}[ht!]
\begin{framed}
\footnotesize

Summary: John Doe focused intensely on security and authentication
aspects in both backend and frontend, implementing crucial features to
ensure system integrity and security.

Contributions:

\begin{itemize}
\item
  \textit{app.py} (Backend): Implemented routes and functionalities
  related to authentication and administration.
\item
  \textit{auth.py}: Developed a complete authentication module.
\item
  \textit{dashboard.py} (Backend): Improved data handling and
  organization.
\item
  \textit{mongo\_users.py}: Implemented functions for recovery code
  updates and password changes.
\item
  \textit{rec\_password.py}: Added update commands in MongoDB.
\item
  \textit{secrets\_.py}: Added a password handling function.
\item
  \textit{pages/cdg\_rec.py} (Frontend): Implemented the code
  verification functionality for password recovery.
\item
  \textit{pages/login.py} (Frontend): Implemented a login form using
  Streamlit.
\item
  \textit{pages/new\_market.py} (Frontend): Developed an interactive
  interface for market creation.
\item
  \textit{pages/rec\_password.py} (Frontend): Developed a password recovery
  system.
\end{itemize}

\textbf{Overall contribution of the team}

This stage of the sprint was crucial for the project's advancement. Through the detailed contributions above, we achieved the following progress:
\begin{itemize}
\item  Development of a robust web application using Flask, with authentication features, user management, and data validation.
\item  Implementation of services that ensure data security, such as password hashing and input validation, essential for application integrity.
\item Creation of an email sending system, crucial for communication within the application context, allowing users to receive relevant information.
\item Establishment of a unit testing foundation, ensuring that application functionalities are maintained and operate correctly during continuous development.
\end{itemize}
\end{framed}
\caption{Summary produced by the tool: contribution of a single student and overall contribution of the team}
\label{summary-eng}
\end{table}

\section{Results and Discussions}

This procedure was tested on a class of 28 students in the first semester of 2024 and a class of 37 students in the second semester of 2024. The summary was run and presented to the students twice in each class during the 3-week sprint (at the end of the second week and at the end of the third week). In total, 130 summaries of student contributions were provided, and there was general agreement among the students that the summaries accurately reflected the tasks they had performed.

There were some errors, including eight instances of partial omissions of student contributions, where the summary included several actual tasks the student had completed but missed others. 
Additionally, there were four cases of total omissions of student contributions. In these instances, students had completed tasks that were not included in the report.

One case of factual inaccuracy occurred when a certain feature was attributed to a student based on a comment in the source code. However, another student had made the commits, but the LLM was led by the comment. This may establish a category of 'comment injections', where a student writes an untrue comment in the code that the LLM accepts as accurate, should LLMs become more widespread for source code analysis.

Excluded from these poor performance measures are cases that were reported by students, but were the default behavior of our system. For instance, branches that were never merged into the main branch were not considered in the summary, and students who did not commit any code during the periods were also omitted from the summary —- four cases.

\subsection{Point of view of the students}

When the feedback was presented, students were invited to volunteer for interviews regarding the LLM-generated summary and other aspects of teamwork. The interviews were not conducted by any course instructors, and students were assured that their responses would not impact their grades. Out of 65 students, 8 volunteered, and the interviews were conducted in 30-minute sessions. Students were asked to review the summary and identify any inaccuracies. They were also questioned about the usefulness of the summary and its effect on team organization.

Overall, students felt that the summary accurately listed the tasks performed. However, their main criticism was that the summary was superficial, failing to recognize that a series of file changes could represent the implementation of an entire feature. Additionally, students noted that the summary tended to inflate the importance of minor tasks, such as writing a README file, using a tool to generate scaffold code or making small code changes. This was a common complaint among students and was also observed by instructors.

Following is an example of what students positively said in the interview about the feedback:

\begin{quote}
"...it's (feedback) very good at seeing many things that people did and reporting that..."
\end{quote}

Another student highlighted that he negatively thinks about the feedback:

\begin{quote}
"I felt that I had done more than what was written. I made the password change function and all that, maybe in the final print it didn't appear, but I worked a lot."
\end{quote}

Students found their assignments to roles based on their contributions to be partially arbitrary or incorrect, making comparisons between students with the same role difficult. Another recurring observation from multiple interviews was to consider co-authorship tags in GitHub, as contributions from pair-programming sessions were sometimes omitted. They appreciated having a summary of the team's work, as it helped organize tasks. An external summary facilitated fairer discussions about task distribution. One student mentioned that the system helped them conduct a difficult conversation when a colleague was assigned the role of "junior documenter", primarily committing comment lines without implementing new code. This allowed the team to assess the individual's limited contribution, motivating improvement in the next sprint.

\begin{quote}
"...but it didn't make much sense when it said 'ah, it's junior, it's senior', and all that. Just because they did some things a little beyond. Because theoretically, it's something we've already learned there, so it's kind of... this question of senior, junior, I think it depends on each company and each place, what is really a junior, what is a senior, you know? So I really didn't find it that viable."
\end{quote}

\subsection{Point of view of the instructors}

The instructors, who are not authors of this paper, but were working directed as supervisors of students during project development, reviewed the summaries and provided feedback on their usefulness and impact on team organization. They found the information consistent with their interactions with the teams, confirming that the tasks accurately reflected student activities.

However, they noted that the tool tended to overvalue minor or automated tasks. It would credit code that was not necessarily written by the students, as code automatically generated by the tools in use, or small README updates, which sometimes misrepresented the actual contributions of some students. This issue was also highlighted by the students.

Instructors found the tool beneficial for identifying students outliers who were not genuinely contributing. While they agreed that it is useful for monitoring teams, it has not yet prompted specific actions related to team management, as it primarily confirmed their existing impressions.

Additionally, the instructors observed that students with previously low participation increased their involvement after the initial feedback. Some students who volunteered for more in-depth interviews reported feeling recognized, which motivated them to improve their attendance and participation.

\section{Limitation}

Although currently in use, the tool remains highly experimental. It was primarily tested on Python projects that included frontend, backend, and often data science components, where the summaries proved useful. Those aspects constitute threats to external validity, as the tool may not work reasonably well in other contexts.

All data regarding the accuracy of the summaries, whether they contained inaccuracies or omitted information, was collected during sessions when the summaries were presented to student groups. Groups received both printed and electronic copies of the summaries and were instructed to review them for inaccuracies. This process took place during a 120-minute session. The student acting as Scrum master was responsible for collecting feedback and reporting it to the instructor, who then documented it. However, the thoroughness of the students' reviews is uncertain, and some inaccuracies may have been overlooked if their evaluations were not comprehensive. This is a threat to internal validity.  

\section{Conclusions}

The total cost for each semester, covering all the generated feedback, amounted to only \$4. This low cost allows the system to be used more frequently or include more evaluations.

As for the questions, we consider that the response to \textbf{Q1} is partially yes. The tool reports, in large proportion, activities that were in fact performed by the students. Failure situations have been discussed.  The \textbf{Q2} was answered positively, and the summaries were considered by the instructors as a useful potential tool to keep up with the projects. The positive aspects were that the tool enabled faculty to be aware of each student's activities concisely; The repeated feedback sessions provided instructors with insights into students' progress over time; The external summary of contributions facilitated social regulation within student teams, allowing groups to discuss and redistribute tasks more fairly without placing the burden of evidence production on individual members. 

Points considered negative are that in the way the processing of information and LLM prompts were structured, even small or irrelevant contributions tended to be inflated. Students who did relevant work felt partially disadvantaged for that. The zero-shot classification in roles made by the LLM sparked curiosity in the teams, and members were eager to see how the role the LLM had put them into, but in a further moment, students got confused trying to compare themselves with colleagues. But there were cases where roles were not comparable across groups, for instance, a student who was classified as a Junior Frontend Developer in a productive group had done more than a Senior Frontend Developer in a less productive group. Due to this and other factors, the classification into roles was considered confusing.

\bibliographystyle{IEEEtran}
\bibliography{sample-base}

\begin{thebibliography}{10}
\providecommand{\url}[1]{#1}
\csname url@samestyle\endcsname
\providecommand{\newblock}{\relax}
\providecommand{\bibinfo}[2]{#2}
\providecommand{\BIBentrySTDinterwordspacing}{\spaceskip=0pt\relax}
\providecommand{\BIBentryALTinterwordstretchfactor}{4}
\providecommand{\BIBentryALTinterwordspacing}{\spaceskip=\fontdimen2\font plus
\BIBentryALTinterwordstretchfactor\fontdimen3\font minus \fontdimen4\font\relax}
\providecommand{\BIBforeignlanguage}[2]{{%
\expandafter\ifx\csname l@#1\endcsname\relax
\typeout{** WARNING: IEEEtran.bst: No hyphenation pattern has been}%
\typeout{** loaded for the language `#1'. Using the pattern for}%
\typeout{** the default language instead.}%
\else
\language=\csname l@#1\endcsname
\fi
#2}}
\providecommand{\BIBdecl}{\relax}
\BIBdecl

\bibitem{lo_what_2023}
\BIBentryALTinterwordspacing
C.~K. Lo, ``\BIBforeignlanguage{en}{What {Is} the {Impact} of {ChatGPT} on {Education}? {A} {Rapid} {Review} of the {Literature}},'' \emph{\BIBforeignlanguage{en}{Education Sciences}}, vol.~13, no.~4, p. 410, Apr. 2023, number: 4 Publisher: Multidisciplinary Digital Publishing Institute. [Online]. Available: \url{https://www.mdpi.com/2227-7102/13/4/410}
\BIBentrySTDinterwordspacing

\bibitem{feldt_ways_2018}
\BIBentryALTinterwordspacing
R.~Feldt, F.~G. de~Oliveira~Neto, and R.~Torkar, ``Ways of applying artificial intelligence in software engineering,'' in \emph{Proceedings of the 6th {International} {Workshop} on {Realizing} {Artificial} {Intelligence} {Synergies} in {Software} {Engineering}}, ser. {RAISE} '18.\hskip 1em plus 0.5em minus 0.4em\relax New York, NY, USA: Association for Computing Machinery, 2018, pp. 35--41. [Online]. Available: \url{https://doi.org/10.1145/3194104.3194109}
\BIBentrySTDinterwordspacing

\bibitem{phung_generating_2023}
\BIBentryALTinterwordspacing
T.~Phung, J.~Cambronero, S.~Gulwani, T.~Kohn, R.~Majumdar, A.~Singla, and G.~Soares, ``Generating {High}-{Precision} {Feedback} for {Programming} {Syntax} {Errors} using {Large} {Language} {Models},'' Apr. 2023, arXiv:2302.04662 [cs]. [Online]. Available: \url{http://arxiv.org/abs/2302.04662}
\BIBentrySTDinterwordspacing

\bibitem{kasneci_chatgpt_2023}
\BIBentryALTinterwordspacing
E.~Kasneci, K.~Sessler, S.~Küchemann, M.~Bannert, D.~Dementieva, F.~Fischer, U.~Gasser, G.~Groh, S.~Günnemann, E.~Hüllermeier, S.~Krusche, G.~Kutyniok, T.~Michaeli, C.~Nerdel, J.~Pfeffer, O.~Poquet, M.~Sailer, A.~Schmidt, T.~Seidel, M.~Stadler, J.~Weller, J.~Kuhn, and G.~Kasneci, ``{ChatGPT} for good? {On} opportunities and challenges of large language models for education,'' \emph{Learning and Individual Differences}, vol. 103, p. 102274, Apr. 2023. [Online]. Available: \url{https://www.sciencedirect.com/science/article/pii/S1041608023000195}
\BIBentrySTDinterwordspacing

\bibitem{baidoo-anu_education_2023}
\BIBentryALTinterwordspacing
D.~Baidoo-Anu and L.~Owusu~Ansah, ``\BIBforeignlanguage{en}{Education in the {Era} of {Generative} {Artificial} {Intelligence} ({AI}): {Understanding} the {Potential} {Benefits} of {ChatGPT} in {Promoting} {Teaching} and {Learning}},'' Rochester, NY, Jan. 2023. [Online]. Available: \url{https://papers.ssrn.com/abstract=4337484}
\BIBentrySTDinterwordspacing

\bibitem{gilson_how_2023}
\BIBentryALTinterwordspacing
A.~Gilson, C.~W. Safranek, T.~Huang, V.~Socrates, L.~Chi, R.~A. Taylor, and D.~Chartash, ``\BIBforeignlanguage{EN}{How {Does} {ChatGPT} {Perform} on the {United} {States} {Medical} {Licensing} {Examination} ({USMLE})? {The} {Implications} of {Large} {Language} {Models} for {Medical} {Education} and {Knowledge} {Assessment}},'' \emph{\BIBforeignlanguage{EN}{JMIR Medical Education}}, vol.~9, no.~1, p. e45312, Feb. 2023, company: JMIR Medical Education Distributor: JMIR Medical Education Institution: JMIR Medical Education Label: JMIR Medical Education Publisher: JMIR Publications Inc., Toronto, Canada. [Online]. Available: \url{https://mededu.jmir.org/2023/1/e45312}
\BIBentrySTDinterwordspacing

\bibitem{dehbozorgi_llm-based_2024}
\BIBentryALTinterwordspacing
N.~Dehbozorgi and M.~T. Kunuku, ``An {LLM}-based {Reflection} {Analysis} {Tool} for {Identifying} and {Addressing} {Challenging} {Topics},'' in \emph{Proceedings of the 55th {ACM} {Technical} {Symposium} on {Computer} {Science} {Education} {V}. 2}, ser. {SIGCSE} 2024.\hskip 1em plus 0.5em minus 0.4em\relax New York, NY, USA: Association for Computing Machinery, Mar. 2024, pp. 1618--1619. [Online]. Available: \url{https://dl.acm.org/doi/10.1145/3626253.3635601}
\BIBentrySTDinterwordspacing

\bibitem{rudolph_chatgpt_2023}
\BIBentryALTinterwordspacing
J.~Rudolph, S.~Tan, and S.~Tan, ``\BIBforeignlanguage{en}{{ChatGPT}: {Bullshit} spewer or the end of traditional assessments in higher education?}'' \emph{\BIBforeignlanguage{en}{Journal of Applied Learning and Teaching}}, vol.~6, no.~1, pp. 342--363, Jan. 2023, number: 1. [Online]. Available: \url{https://journals.sfu.ca/jalt/index.php/jalt/article/view/689}
\BIBentrySTDinterwordspacing

\bibitem{cope_artificial_2021}
\BIBentryALTinterwordspacing
B.~Cope, M.~Kalantzis, and D.~Searsmith, ``Artificial intelligence for education: {Knowledge} and its assessment in {AI}-enabled learning ecologies,'' \emph{Educational Philosophy and Theory}, vol.~53, no.~12, pp. 1229--1245, Oct. 2021, publisher: Routledge \_eprint: https://doi.org/10.1080/00131857.2020.1728732. [Online]. Available: \url{https://doi.org/10.1080/00131857.2020.1728732}
\BIBentrySTDinterwordspacing

\bibitem{kirova_software_2024}
\BIBentryALTinterwordspacing
V.~D. Kirova, C.~S. Ku, J.~R. Laracy, and T.~J. Marlowe, ``\BIBforeignlanguage{en}{Software {Engineering} {Education} {Must} {Adapt} and {Evolve} for an {LLM} {Environment}},'' in \emph{\BIBforeignlanguage{en}{Proceedings of the 55th {ACM} {Technical} {Symposium} on {Computer} {Science} {Education} {V}. 1}}.\hskip 1em plus 0.5em minus 0.4em\relax Portland OR USA: ACM, Mar. 2024, pp. 666--672. [Online]. Available: \url{https://dl.acm.org/doi/10.1145/3626252.3630927}
\BIBentrySTDinterwordspacing

\bibitem{robles_development_2022}
\BIBentryALTinterwordspacing
G.~Robles, A.~Capiluppi, J.~M. Gonzalez-Barahona, B.~Lundell, and J.~Gamalielsson, ``\BIBforeignlanguage{en}{Development effort estimation in free/open source software from activity in version control systems},'' \emph{\BIBforeignlanguage{en}{Empirical Software Engineering}}, vol.~27, no.~6, p. 135, Jul. 2022, number: 6. [Online]. Available: \url{https://doi.org/10.1007/s10664-022-10166-x}
\BIBentrySTDinterwordspacing

\bibitem{radermacher_investigating_2014}
\BIBentryALTinterwordspacing
A.~Radermacher, G.~Walia, and D.~Knudson, ``Investigating the skill gap between graduating students and industry expectations,'' in \emph{Companion {Proceedings} of the 36th {International} {Conference} on {Software} {Engineering}}, ser. {ICSE} {Companion} 2014.\hskip 1em plus 0.5em minus 0.4em\relax New York, NY, USA: Association for Computing Machinery, 2014, pp. 291--300. [Online]. Available: \url{https://dl.acm.org/doi/10.1145/2591062.2591159}
\BIBentrySTDinterwordspacing

\bibitem{ruf_communication_2009}
\BIBentryALTinterwordspacing
S.~Ruf and M.~Carter, ``Communication learning outcomes from software engineering professionals: {A} basis for teaching communication in the engineering curriculum,'' \emph{IEEE}, Oct. 2009, accepted: 2010-12-09T19:44:39Z ISBN: 9781424447152 Publisher: Institute of Electrical and Electronics Engineers. [Online]. Available: \url{https://dspace.mit.edu/handle/1721.1/60260}
\BIBentrySTDinterwordspacing

\bibitem{ettington_facilitating_2002}
\BIBentryALTinterwordspacing
D.~R. Ettington and R.~R. Camp, ``\BIBforeignlanguage{en}{Facilitating {Transfer} of {Skills} between {Group} {Projects} and {Work} {Teams}},'' \emph{\BIBforeignlanguage{en}{Journal of Management Education}}, vol.~26, no.~4, pp. 356--379, Aug. 2002, publisher: SAGE Publications Inc. [Online]. Available: \url{https://doi.org/10.1177/105256290202600404}
\BIBentrySTDinterwordspacing

\bibitem{bacon_lessons_1999}
\BIBentryALTinterwordspacing
D.~R. Bacon, K.~A. Stewart, and W.~S. Silver, ``\BIBforeignlanguage{en}{Lessons from the {Best} and {Worst} {Student} {Team} {Experiences}: {How} a {Teacher} can make the {Difference}},'' \emph{\BIBforeignlanguage{en}{Journal of Management Education}}, vol.~23, no.~5, pp. 467--488, Oct. 1999, publisher: SAGE Publications Inc. [Online]. Available: \url{https://doi.org/10.1177/105256299902300503}
\BIBentrySTDinterwordspacing

\bibitem{feichtner_why_1984}
\BIBentryALTinterwordspacing
S.~B. Feichtner and E.~A. Davis, ``\BIBforeignlanguage{en}{Why {Some} {Groups} {Fail}: a {Survey} of {Students}' {Experiences} with {Learning} {Groups}},'' \emph{\BIBforeignlanguage{en}{Organizational Behavior Teaching Review}}, vol.~9, no.~4, pp. 58--73, Nov. 1984, publisher: SAGE Publications. [Online]. Available: \url{https://journals.sagepub.com/doi/abs/10.1177/105256298400900409}
\BIBentrySTDinterwordspacing

\bibitem{falkner_collaborative_2013}
\BIBentryALTinterwordspacing
K.~Falkner, N.~J. Falkner, and R.~Vivian, ``Collaborative learning and anxiety: a phenomenographic study of collaborative learning activities,'' in \emph{Proceeding of the 44th {ACM} technical symposium on {Computer} science education}, ser. {SIGCSE} '13.\hskip 1em plus 0.5em minus 0.4em\relax New York, NY, USA: Association for Computing Machinery, Mar. 2013, pp. 227--232. [Online]. Available: \url{https://dl.acm.org/doi/10.1145/2445196.2445268}
\BIBentrySTDinterwordspacing

\bibitem{ohland_comprehensive_2012-1}
\BIBentryALTinterwordspacing
M.~W. Ohland, M.~L. Loughry, D.~J. Woehr, L.~G. Bullard, R.~M. Felder, C.~J. Finelli, R.~A. Layton, H.~R. Pomeranz, and D.~G. Schmucker, ``The {Comprehensive} {Assessment} of {Team} {Member} {Effectiveness}: {Development} of a {Behaviorally} {Anchored} {Rating} {Scale} for {Self}- and {Peer} {Evaluation},'' \emph{Academy of Management Learning \& Education}, vol.~11, no.~4, pp. 609--630, Dec. 2012, publisher: Academy of Management. [Online]. Available: \url{https://journals.aom.org/doi/10.5465/amle.2010.0177}
\BIBentrySTDinterwordspacing

\bibitem{willcoxson_its_2006}
\BIBentryALTinterwordspacing
L.~E. Willcoxson, ``\BIBforeignlanguage{en}{“{It}’s not {Fair}!”: {Assessing} the {Dynamics} and {Resourcing} of {Teamwork}},'' \emph{\BIBforeignlanguage{en}{Journal of Management Education}}, vol.~30, no.~6, pp. 798--808, Dec. 2006, publisher: SAGE Publications Inc. [Online]. Available: \url{https://doi.org/10.1177/1052562906287964}
\BIBentrySTDinterwordspacing

\bibitem{tarmazdi_using_2015}
\BIBentryALTinterwordspacing
H.~Tarmazdi, R.~Vivian, C.~Szabo, K.~Falkner, and N.~Falkner, ``Using {Learning} {Analytics} to {Visualise} {Computer} {Science} {Teamwork},'' in \emph{Proceedings of the 2015 {ACM} {Conference} on {Innovation} and {Technology} in {Computer} {Science} {Education}}, ser. {ITiCSE} '15.\hskip 1em plus 0.5em minus 0.4em\relax New York, NY, USA: Association for Computing Machinery, Jun. 2015, pp. 165--170. [Online]. Available: \url{https://dl.acm.org/doi/10.1145/2729094.2742613}
\BIBentrySTDinterwordspacing

\bibitem{sandee_gitcanary_2020}
\BIBentryALTinterwordspacing
J.~J. Sandee and E.~Aivaloglou, ``{GitCanary}: {A} {Tool} for {Analyzing} {Student} {Contributions} in {Group} {Programming} {Assignments},'' in \emph{Proceedings of the 20th {Koli} {Calling} {International} {Conference} on {Computing} {Education} {Research}}, ser. Koli {Calling} '20.\hskip 1em plus 0.5em minus 0.4em\relax New York, NY, USA: Association for Computing Machinery, Nov. 2020, pp. 1--2. [Online]. Available: \url{https://dl.acm.org/doi/10.1145/3428029.3428563}
\BIBentrySTDinterwordspacing

\bibitem{gousios_measuring_2008}
\BIBentryALTinterwordspacing
G.~Gousios, E.~Kalliamvakou, and D.~Spinellis, ``\BIBforeignlanguage{en}{Measuring developer contribution from software repository data},'' in \emph{\BIBforeignlanguage{en}{Proceedings of the 2008 international working conference on {Mining} software repositories}}.\hskip 1em plus 0.5em minus 0.4em\relax Leipzig Germany: ACM, May 2008, pp. 129--132. [Online]. Available: \url{https://dl.acm.org/doi/10.1145/1370750.1370781}
\BIBentrySTDinterwordspacing

\bibitem{lima_assessing_2015}
\BIBentryALTinterwordspacing
J.~Lima, C.~Treude, F.~F. Filho, and U.~Kulesza, ``Assessing developer contribution with repository mining-based metrics,'' in \emph{2015 {IEEE} {International} {Conference} on {Software} {Maintenance} and {Evolution} ({ICSME})}, Sep. 2015, pp. 536--540. [Online]. Available: \url{https://ieeexplore.ieee.org/document/7332509}
\BIBentrySTDinterwordspacing

\bibitem{buffardi_assessing_2020}
\BIBentryALTinterwordspacing
K.~Buffardi, ``Assessing {Individual} {Contributions} to {Software} {Engineering} {Projects} with {Git} {Logs} and {User} {Stories},'' in \emph{Proceedings of the 51st {ACM} {Technical} {Symposium} on {Computer} {Science} {Education}}, ser. {SIGCSE} '20.\hskip 1em plus 0.5em minus 0.4em\relax New York, NY, USA: Association for Computing Machinery, 2020, pp. 650--656. [Online]. Available: \url{https://doi.org/10.1145/3328778.3366948}
\BIBentrySTDinterwordspacing

\bibitem{hundhausen_evaluating_2021-1}
\BIBentryALTinterwordspacing
C.~Hundhausen, A.~Carter, P.~Conrad, A.~Tariq, and O.~Adesope, ``Evaluating {Commit}, {Issue} and {Product} {Quality} in {Team} {Software} {Development} {Projects},'' in \emph{Proceedings of the 52nd {ACM} {Technical} {Symposium} on {Computer} {Science} {Education}}, ser. {SIGCSE} '21.\hskip 1em plus 0.5em minus 0.4em\relax New York, NY, USA: Association for Computing Machinery, Mar. 2021, pp. 108--114. [Online]. Available: \url{https://dl.acm.org/doi/10.1145/3408877.3432362}
\BIBentrySTDinterwordspacing

\bibitem{igor_2024}
\BIBentryALTinterwordspacing
I.~dos Santos~Montagner, R.~Corsi Ferr\~{a}o, A.~Kurauchi, M.~Silva, and C.~Zilles, ``Evaluating mastery-oriented grading in an intensive cs1 course,'' in \emph{Proceedings of the 55th ACM Technical Symposium on Computer Science Education V. 1}, ser. SIGCSE 2024.\hskip 1em plus 0.5em minus 0.4em\relax New York, NY, USA: Association for Computing Machinery, 2024, p. 303–309. [Online]. Available: \url{https://doi-org.proxy2.library.illinois.edu/10.1145/3626252.3630841}
\BIBentrySTDinterwordspacing

\bibitem{cliquet_developing_2023}
\BIBentryALTinterwordspacing
G.~Cliquet, F.~Miranda, G.~Tonin, and G.~Rodrigues, ``Developing {Teamwork} {With} the {Aid} of {Reflection},'' São Paulo, 2023. [Online]. Available: \url{https://www.insper.edu.br/en/paee-ale-2023/proceedings/PAEE_ALE_2023_PROCEEDINGS_pre_print.pdf}
\BIBentrySTDinterwordspacing

\bibitem{lin_effects_2018}
\BIBentryALTinterwordspacing
J.-W. Lin, ``Effects of an online team project-based learning environment with group awareness and peer evaluation on socially shared regulation of learning and self-regulated learning,'' \emph{Behaviour \& Information Technology}, vol.~37, no.~5, pp. 445--461, May 2018, publisher: Taylor \& Francis \_eprint: https://doi.org/10.1080/0144929X.2018.1451558. [Online]. Available: \url{https://doi.org/10.1080/0144929X.2018.1451558}
\BIBentrySTDinterwordspacing

\bibitem{miranda_2024}
\BIBentryALTinterwordspacing
F.~de~Miranda, R.~C. Ferrao, D.~P. Soler, and M.~A. Vieira~Graglia, ``Llm-based individual contribution summarization in software projects,'' in \emph{Proceedings of the 2024 on ACM Virtual Global Computing Education Conference V. 2}, ser. SIGCSE Virtual 2024.\hskip 1em plus 0.5em minus 0.4em\relax New York, NY, USA: Association for Computing Machinery, 2024, p. 307–308. [Online]. Available: \url{https://doi-org.proxy2.library.illinois.edu/10.1145/3649409.3691092}
\BIBentrySTDinterwordspacing

\bibitem{spadini2018pydriller}
\BIBentryALTinterwordspacing
D.~Spadini, M.~Aniche, and A.~Bacchelli, ``Pydriller: Python framework for mining software repositories,'' in \emph{Proceedings of the 2018 26th ACM Joint Meeting on European Software Engineering Conference and Symposium on the Foundations of Software Engineering}, ser. ESEC/FSE 2018.\hskip 1em plus 0.5em minus 0.4em\relax New York, NY, USA: Association for Computing Machinery, 2018, p. 908–911. [Online]. Available: \url{https://doi-org.proxy2.library.illinois.edu/10.1145/3236024.3264598}
\BIBentrySTDinterwordspacing

\end{thebibliography}

\end{document}